\documentclass[aps,pra,twocolumn,10pt,nobalancelastpage,altaffilletter,superscriptaddress,longbibliography]{revtex4-2}

\usepackage[utf8x]{inputenc}
\usepackage[T1]{fontenc}
\usepackage{amsmath, amssymb}
\usepackage{bm}
    \renewcommand{\vec}{\bm}

\usepackage{graphicx}
\usepackage{braket}
\usepackage{lipsum}

\usepackage[breaklinks, pdftex, hyperfootnotes=true, pdfpagelabels, bookmarks, pageanchor]{hyperref}

\hypersetup{colorlinks=true, pageanchor=true, urlcolor=blue, linkcolor=blue, citecolor=blue}

\usepackage[normalem]{ulem}
\newcommand{\add}[1]{#1}
\newcommand{\del}[1]{}
\newcommand{\rep}[2]{\del{#1}{}\add{#2}}

\DeclareMathOperator{\Tr}{Tr}

\begin{document}

\title{Nonstabilizerness of a Boundary Time Crystal}
\author{Gianluca Passarelli}
\email{gianluca.passarelli@unina.it}
\author{Angelo Russomanno}
\author{Procolo Lucignano}
\affiliation{Dipartimento di Fisica, Universit\`a di Napoli ``Federico II'', I-80126 Napoli, Italy}

\begin{abstract}
    Boundary time crystals exhibit measurement-induced phase transitions in their steady-state entanglement, with critical behavior that depends on the particular unraveling of the Lindblad dynamics. In this work, we investigate another key measure of quantum complexity---nonstabilizerness (or ``magic'')---and show that it follows a markedly different pattern. Importantly, in contrast to entanglement, for large system sizes, \rep{magic}{nonstabilizerness} remains invariant under different unraveling schemes---a property we attribute to the inherent permutational symmetry of the model. Although the steady-state stabilizer entropy does not display a genuine phase transition, it exhibits a singular derivative (a cusp) at the mean-field critical point.  Furthermore, we demonstrate that finite-size simulations of the average Lindblad evolution fail to capture the asymptotic behavior of \rep{magic}{nonstabilizerness} in the time-crystal phase, while quantum trajectory unravelings correctly reveal its extensive scaling with system size. These findings offer insights into how different quantum resources manifest in open systems.
\end{abstract}

\maketitle

\section{Introduction}
\label{sec:intro}

Collective spin models describe ensembles of two-level systems, or qubits, in which all particles experience the same global dynamics due to collective interactions, leading to permutationally invariant quantum states. These models are particularly relevant in quantum optics, where atoms confined in a cavity interact through a shared electromagnetic field~\cite{walls1978nonequilibriumcooperative,puri1979steadystatecollective,walls1980cooperativefluorescence,schneider2002entanglementsteadystatecollective,chase2008,hannukainen2018dissipationdrivenqpt,shammah2018}. A prominent example is the boundary time crystal (BTC), a driven-dissipative many-body system that exhibits persistent oscillations, breaking \add{continuous} time-translation symmetry~\cite{iemini2018}. BTCs arise from the interplay between coherent driving and collective dissipation and have been proposed as promising platforms for investigating nonequilibrium quantum phenomena, leading to a flourishing research field~\cite{tucker2018shatteredtime,zhu2019dicketc,lledo2019drivenbosehubbard,prazeres2021btcdlevel,piccitto2021symmetriesbtc,carollo2021exactnessmf,passarelli2022dissipativetimecrystals,lourenco2022multipartitebtc,buca2022quantumsynch,seetharam2022correlationengineering,carollo2022exactsolutionbtc,hajdusfiek2022seeding,cabot2023trajectoriesbtc,carollo2023,ferioli2023nature,montenegro2023metrologybtc,souza2023gaplesslindbladian,carollo2024sensingbtc,cabot2024nonequilibrium,barberena2024criticalsteadystates,dutta2025limitcycles}. In particular, they are known to host measurement-induced phase transitions (MIPTs)~\cite{skinner2019,cao2019freefermions,szyniszewski2020universalitymipt,buchhold2021effectivemipt,alberton2021freefermions,potter2022,sierant2022}, where their entanglement exhibits a transition as a function of the dissipation rate, or measurement strength~\cite{li2019,szyniszewski2019entanglementtransition,chaoming2020measurementinducedcriticality,fuji2020measurementinducedcriticality,turkeshi2020measurementindicedcriticality,zabalo2020miptrandomcircuits,lang2020tfi,lunt2020miptmbl,nahum2021miptalltoall,turkeshi2021,ippoliti2021measurementonly,sierant2022-2,altland2022,coppola2022growthentanglement,minato2022miptlongrange,passarelli2023,piccitto2023stringmeasurement,delmonte2024,lirasolanilla2024multipartiteentanglement,chahine2024entanglementfreefermions,legal2024entanglementdynamics}. This naturally raises the question of whether other quantum complexity measures, beyond entanglement, might also undergo phase transitions in monitored BTCs.

Nonstabilizerness, one such measure, quantifies the deviation of a quantum state from the set of stabilizers~\cite{bravyi2005,howard2014,howard2014-2,bravyi2016,bravyi2016-2,bravyi2019,heinrich2019,wang2019,wang2020,zhou2020,heimendahl2021,jiang2023}. This concept is crucial in quantum computing because stabilizer states, which can be efficiently prepared using Clifford gates, are classically simulable and do not offer a computational advantage~\cite{gottesman1997,gottesman1998,aaronson2004,fisher2023}. While Clifford gates---the Hadamard (\textsf{H}), Phase (\textsf{S}), and controlled-NOT (\textsf{CNOT}) gates---are fundamental for quantum error correction and can be efficiently simulated via the Gottesman-Knill theorem~\cite{gottesman1998-2}, they alone do not enable universal quantum computation. To surpass classical capabilities, nonstabilizer or \textit{magic} states must be introduced, typically through \textsf{T}-gates~\cite{nielsen2010}. Consequently, nonstabilizerness serves as a key measure of the quantum computational power of many-body systems~\cite{seddon2019,rattacaso2023,leone2023nonstabilizernesshardnessfidelity,dowling2024magicheisenbergpicture}.

While the role of entanglement in characterizing quantum phases and dynamics is well established~\cite{amico2008,laflorencie2016,lunt2022,passarelli2023}, nonstabilizerness in many-body physics remains less explored~\cite{liu2022}, especially in the context of phase transitions, where only a handful of results are known~\cite{oliviero2022,leone2023,niroula2023,fux2023,bejan2023,tarabunga2023-3,tarabunga2024magictransitionmeasurementonlycircuits}. Investigating nonstabilizerness in monitored many-body systems, such as BTCs, can provide insights into quantum complexity in open dynamics and potentially reveal new universal features of measurement-induced transitions. In particular, understanding whether \rep{magic}{nonstabilizerness} exhibits a transition similar to that of entanglement in monitored BTCs could shed light on how different forms of quantum complexity behave under nonunitary evolution.

Motivated by this, our goal is to study the monitored dynamics of a model admitting a boundary time crystal phase, i.\,e., a permutationally invariant system composed of $ N $ qubits placed in an optical cavity, driven by an external laser field, and subject to collective Markovian \add{(memoryless)} decay~\cite{iemini2018,hannukainen2018dissipationdrivenqpt,carollo2023,ferioli2023nature}. The unconditional dynamics of the system (i.\,e., the average over many independent experiments) is described by a master equation in the Lindblad form, which is able to capture a superradiant transition in the steady state\add{, where transition processes are described by the action of a ``jump operator'' acting on the system state}. \rep{However, w}{W}hen coupled to a suitable \del{measuring} device that monitors and keeps track of environmental transitions induced by the \rep{jump operator}{measurements} (i.\,e., the evolution \emph{conditioned} to the measurement outcomes), the dynamics can be described in terms of stochastic quantum trajectories~\cite{carmichael1999statisticalmethodsin,jacobs2014quantummeasurementtheory} that, \add{only} on average, provide the Lindblad equation. Depending on the experimental apparatus, different measurement schemes---ranging from photodetection\add{, where one counts the number of photons escaping the cavity,} to homo/heterodyne detection\add{, where one instead measures the quadratures of the outgoing light field by interfering it with a reference signal}---give rise to distinct unravelings of the same Lindbladian. While the evolution of any observable remains identical across different unravelings, nonlinear quantities such as entanglement and \rep{magic}{nonstabilizerness} may depend on the specific measurement protocol~\cite{wiseman2009quantummeasurementand,piccitto2022differentunravellings,Piccitto_2024,passarelli2023,PhysRevB.106.219901}.

This superradiant model presents several advantages for investigating the monitored dynamics of \rep{magic}{nonstabilizerness}. First, the reconstruction of quantum trajectories from experiments is feasible, as individual trajectories saturate within a logarithmic timescale in system size, thereby reducing postselection requirements~\cite{passarelli2023}. Second, the model exhibits a measurement-induced phase transition in entanglement, which coincides with the superradiant phase transition in the average state~\cite{iemini2018,hannukainen2018dissipationdrivenqpt,carollo2023,passarelli2023}. This naturally raises the question of whether \rep{magic}{nonstabilizerness} undergoes a similar transition at the same critical point. Third, the symmetries of the model enable the application of efficient computational techniques for quantifying nonstabilizerness. 

Indeed, most existing studies on nonstabilizerness in many-body systems are constrained to small system sizes~\cite{oliviero2022,rattacaso2023,niroula2023,leone2023,russomanno2025syk} or to systems that admit efficient classical representations, such as sign-problem-free Hamiltonians~\cite{liu2025qmc-sre}, tensor networks or Gaussian states~\cite{haug2023,haug2023-2,lami2023,fux2023,tarabunga2024,mello2024hsmps,frau2024nonstabilizernessversusentanglement,tirrito2024anticoncentrationmagicspreadingergodic,collura2025quantummagicfermionicgaussian,turkeshi2025magicspreadingrandomquantum}. Among these, permutationally invariant states---such as those describing boundary time crystal states---constitute a particularly advantageous platform. Their high degree of symmetry allows for the efficient computation of magic measures, particularly those based on the Pauli spectrum \add{(i.\,e., the expectation values of all multi-qubit Pauli operators)}, with polynomial rather than exponential scaling~\cite{passarelli2024magic}. This renders them especially valuable for exploring nonstabilizerness in large systems~\cite{Passarelli:2024lpm}. Finally, despite its growing relevance in the Noisy Intermediate-Scale Quantum (NISQ) era, dominated by noise and environmental interactions, the role of \rep{magic}{nonstabilizerness} in open and monitored quantum systems remains largely unexplored~\cite{fux2023,Passarelli:2024lpm,tarabunga2024magictransitionmeasurementonlycircuits,russomanno2025syk}.

\add{We emphasize that although the nonstabilizerness is a quantum complexity measure important for quantum computation, and existing quantum computers have only few qubits, it is important to study it in a many-body context as we are doing here. In some sense the situation is similar to the one of the entanglement entropy. Also this quantity was introduced as a measure of quantum complexity for quantum computation, and then its scaling with the system size has been found to have a relevant role in quantum many-body physics, in the context of ground-state quantum phase transitions~\cite{amico2008,calabrese2004}, in distinguishing quantum thermalization from quantum integrable behavior~\cite{Alba_2017,Alba_2018,Caleb_SciPostPhysLectNotes20,Singh_2016,serbyn2013universal,bardarson2012unbounded,Montangero,Prosen_PRB08,PhysRevB.102.144302,Abanin_RMP} and in the measurement-induced phase transitions~\cite{li2019,szyniszewski2019entanglementtransition,chaoming2020measurementinducedcriticality,fuji2020measurementinducedcriticality,turkeshi2020measurementindicedcriticality,zabalo2020miptrandomcircuits,lang2020tfi,lunt2020miptmbl,nahum2021miptalltoall,turkeshi2021,ippoliti2021measurementonly,sierant2022-2,altland2022,coppola2022growthentanglement,minato2022miptlongrange,passarelli2023,piccitto2023stringmeasurement,delmonte2024,lirasolanilla2024multipartiteentanglement,chahine2024entanglementfreefermions,legal2024entanglementdynamics}. In all these cases the scaling of the entanglement entropy with the system size was relevant to highlight behaviors appearing in the large-size limit. For the nonstabilizerness  the situation is still at the beginning, but appears to be similar. For instance this quantity has already been used to characterize ground-state quantum phase transitions~\cite{PhysRevB.110.045101,PhysRevB.111.L081102,viscardi2025interplayentanglementstructuresstabilizer}. Our work provides a further contribution in this direction, focusing on a quantum phase transition involving the full monitored dynamics.}

We first find that evaluating \rep{magic}{nonstabilizerness} across trajectories yields an average that, in the thermodynamic limit, becomes independent of the specific unraveling protocol. This result can be shown analytically using a replica approach and relies on the fact that trajectory correlations become negligible in the thermodynamic limit when computing expectation values. This contrasts sharply with another measure of quantum complexity---the entanglement entropy---which exhibits a significant dependence on the unraveling scheme when averaged over trajectories~\cite{passarelli2023}.

Approaching the thermodynamic limit, we observe that the time average of the trajectory-averaged \rep{magic}{nonstabilizerness} is well captured by mean-field (MF) theory, provided the mean-field results are averaged over time and many random initial conditions. The stochasticity introduced by individual trajectories is effectively described by the average over random initial states in the MF limit. However, this agreement holds only for time averages, not for individual time traces, and manifests as a cusp at the transition to the time-crystal phase. Furthermore, \rep{magic}{nonstabilizerness} scales linearly with the system size $N$ asymptotically, consistent with previous findings in other monitored systems~\cite{russomanno2025syk}. This indicates the absence of \rep{magic}{nonstabilizerness} phase transitions in the model, although \rep{remnants}{signs} of critical behavior are still visible in the cusp.

Interestingly, evaluating \rep{magic}{nonstabilizerness} on the Lindblad-averaged density matrix yields a time average that is not well described by the mean-field theory in the boundary time crystal phase for large system sizes. In this phase, the average density matrix remains mixed and does not approach a pure state in the thermodynamic limit, leading to the failure of mean-field theory in describing \rep{magic}{nonstabilizerness}. This highlights that, while correlations are negligible when computing expectation values of observables, they still affect the mixed state in the large-size limit, as revealed by the behavior of \rep{magic}{nonstabilizerness}.

The rest of this manuscript is organized as follows. In Sec.~\ref{sec:model}, we describe the superradiant model and the monitoring protocols adopted in our analysis. In Sec.~\ref{sec:magic}, we recall the definition of the magic measure considered in this paper, the stabilizer 2-R\'enyi entropy, and briefly discuss how to compute it efficiently in the case of permutationally invariant quantum states. We show also  that the stabilizer entropy is mostly unraveling-independent for large system sizes. In Sec.~\ref{sec:mf}, we discuss the thermodynamic limit of the model, which is exactly captured by mean-field theory. In Sec.~\ref{sec:finite-size}, we turn our attention to finite-size systems and study the dynamics of \rep{magic}{the stabilizer entropy} of the average state and compare the \rep{magic}{nonstabilizerness} averaged over trajectories and time with the mean-field results averaged over time and initial conditions finding a good agreement. There is no agreement with the Lindblad average density matrix, which remains mixed also in the thermodynamic limit, a fact related to the presence of correlations. In Sec.~\ref{sec:conclusions}, we draw our conclusions.

\section{Model and unraveling protocols}
\label{sec:model}

We consider a system of $N$ identical spin-($1/2$) particles with total spin $S = N/2$, whose Hamiltonian is given by the one-body operator,
\begin{equation}\label{eq:hamiltonian-btc}
    \hat H = \omega_0 \hat S_x,
\end{equation}
where $\hat S_\alpha = (1/2) \sum_{k=1}^N \hat \sigma_k^\alpha$ are collective spin operators aligned in the direction $\alpha \in \lbrace x, y, z \rbrace$, and $\omega_0$ is the driving frequency.

We assume that this system is placed in an optical cavity and that the spins interact with the photons of the cavity via a Tavis-Cummings interaction Hamiltonian~\cite{TavisCummings}, $\propto \hat S_- \hat a^\dagger + \hat S_+ \hat a$, where $\hat S_\pm = \hat S_x \pm i \hat S_y$ and $\hat a$ is the bosonic annihilation operator of the cavity mode. The cavity is lossy and the photon emission is expressed by a Lindblad master equation where the jump operator is proportional to the bosonic operator $\hat a$. The emitted photons escape the cavity and are registered by a measurement apparatus (here and in the following we assume perfect detection efficiency), allowing to monitor the combined state of the cavity and atoms. Through a process of adiabatic elimination of the photonic degrees of freedom of the cavity~\cite{mattes2023entangledlightmatter}, this model can be recast in terms of the following Lindblad equation for the spin system alone~\cite{iemini2018},
\begin{equation}\label{eq:btc}
    \partial_t \hat \rho = -i \omega_0 \left[\hat S_x, \hat \rho \right] + \frac{\kappa}{S} \left(\hat S_- \hat \rho \hat S_+ - \frac{1}{2} \left\lbrace \hat S_- \hat S_+, \hat \rho\right\rbrace\right),
\end{equation}
where $\kappa$ is the decay rate and $\hat \rho$ is the (generally mixed) density matrix of the spin cloud. In this picture, the detection of a photon by the monitoring device is associated to the jump operator $\hat L = \sqrt{\kappa / S} \, \hat S_-$: due to the Tavis-Cummings form of the light-matter interaction Hamiltonian, which preserves the total number of excitations in the system, a photon emission is always associated to an atomic decay for the spin system. Thus, the detector is able to continuously track the atomic transitions induced by the interaction with the environment, essentially monitoring the state of the system. Disregarding the measurement outcomes yields the unconditional Lindblad equation~\eqref{eq:btc}. 

If one instead keeps track of the measurement outcomes, it becomes possible to unravel the dissipative dynamics of Eq.~\eqref{eq:btc} in terms of stochastic pure-state quantum trajectories, whose average gives back the state $\hat \rho$. Notice that the Hamiltonian and the jump operators are both expressed in terms of collective spin operators, which commute with the total spin operator $\hat{\vec{S}}^2$. Thus, $\mathrm{SU}(2)$ is a ``strong symmetry'' of this model~\cite{booker-njp-2020:btc-spectral-properties-and-finite-size-effects} and the dynamics entirely takes place in the subspace of the Hilbert space with maximum spin $S = N/2$~\cite{shammah2018}, both for the average state $\hat \rho$ and at the level of individual quantum trajectories. There are infinitely many different unravelings yielding the same Lindblad evolution, which may however result in different ensemble properties of nonlinear quantities of the quantum state~\cite{piccitto2022differentunravellings}.

\subsection{Quantum Jumps}

The quantum jump (QJ) unraveling corresponds to a photodetection measurement scheme, wherein the detector performs projective measurements of the bath in the photon number basis $\{\ket{n_B}\}$, with $\hat a^\dagger \hat a\, \ket{n_B} = n_B \ket{n_B}$~\cite{wiseman2009quantummeasurementand}. Over a small time interval $\delta t$, the system evolves as follows: If no photon is detected, the state evolves deterministically according to the non-Hermitian Hamiltonian
\[
    \hat H_\text{nH} = \hat H - \frac{i}{2} \hat L^\dagger \hat L,
\]
where $\hat L = \sqrt{\kappa/S}\,\hat S_-$.
If a photon is detected (with probability $\delta p = \langle \psi|\hat L^\dagger \hat L|\psi\rangle\, \delta t$), the state undergoes a quantum jump,
\[
    \ket{\psi} \to \frac{\hat L \ket{\psi}}{\sqrt{\langle \psi|\hat L^\dagger \hat L|\psi\rangle}}.
\]
Thus, the update rule to first order in $\delta t$ is~\cite{wiseman2009quantummeasurementand}
\begin{align}
    \delta \ket{\psi} &= \left(-i \hat H_\text{nH}\, \delta t + \frac{\delta p}{2}\right) \ket{\psi} \notag \\
    &\quad+ dN \left(\frac{\hat L}{\sqrt{\langle \psi|\hat L^\dagger \hat L|\psi\rangle}} - \hat{\mathbb{I}}\right) \ket{\psi},
\end{align}
where $dN=1$ with probability $\delta p$ and $dN=0$ with probability $1-\delta p$.

\subsection{Quantum State Diffusion}

An alternative approach is provided by quantum state diffusion (QSD), which is associated with a heterodyne detection scheme wherein the quadratures of the bosonic field are continuously measured. In the QSD formalism, each pure state evolves according to the stochastic differential equation~\cite{wiseman2009quantummeasurementand}
\begin{align}\label{eq:qsd}
    \delta \ket{\psi} &= \Bigl[-i \hat H\, \delta t - \frac{1}{2}\Bigl(\hat L^\dagger \hat L + |\ell|^2 - 2\ell^* \hat L\Bigr)\, \delta t \notag \\
    &\quad\quad + (\hat L - \ell)\, \delta W\Bigr] \ket{\psi},
\end{align}
where $\ell = \langle \psi|\hat L|\psi\rangle$ and $\delta W = \sqrt{\delta t/2}\,(X + iY)$, with $X$ and $Y$ being independent real Gaussian random variables of zero mean and unit variance. Unlike the discrete jumps in the QJ scheme, QSD features a continuous, diffusive evolution of the state.

\subsection{General Unravelings}

The QJ and QSD schemes are two limiting cases of a more general detection protocol in which the system is coupled to a local oscillator with frequency $\mathrm{Re}(\mu)$, where $\mu \in \mathbb{C}$~\footnote{The imaginary part $\mathrm{Im}(\mu)$ is related to phase-sensitive attenuation or amplification in the detection process.}. In this more general framework, the jump operator and the Hamiltonian are shifted as follows~\cite{wiseman2009quantummeasurementand}:
\begin{equation}\label{eq:unravelings}
    \hat L \to \hat L + \mu, \qquad \hat H \to \hat H - \frac{i}{2} \left(\mu^* \hat L - \mu \hat L^\dagger\right).
\end{equation}
Setting $\mu = 0$ recovers the conventional QJ unraveling, while the limit $\mu \to \infty$ leads to a diffusive unraveling akin to QSD. Importantly, for all values of $\mu$, the ensemble-averaged dynamics reproduces the Lindblad evolution of Eq.~\eqref{eq:btc}. However, the statistical properties of nonlinear observables (e.\,g., the stabilizer entropy) may vary with the choice of unraveling, thereby offering a versatile tool to probe the interplay between measurement backaction and quantum complexity.

The flexibility in the choice of unraveling protocol enables a detailed investigation into how measurement schemes and the associated stochastic dynamics affect the quantum properties of the system. While linear observables remain invariant across different unravelings, nonlinear measures such as entanglement and \rep{magic}{nonstabilizerness} can exhibit different behaviors under different unraveling protocols, providing deeper insights into the role of monitoring in open quantum systems.

\section{Stabilizer entropy}\label{sec:magic}

Stabilizer entropies are commonly used to quantify \rep{quantum magic}{nonstabilizerness} and are magic monotones for R\'enyi indices $k\ge2$~\cite{haug2023,leone2024monotones}. 
For mixed states, the 2-R\'enyi stabilizer entropy (SRE) is expressed as~\cite{leone2022}
\begin{equation}\label{eq:sre-density}
    \mathcal{M}_2(\hat \rho) = -\log \sum_{\hat P \in \mathcal{P}_N} \frac{\mathrm{Tr}^4(\hat P \hat \rho)}{2^N} + \log \mathrm{Tr}(\hat \rho^2),
\end{equation}
where $\add{\hat P = \sigma_1^{\alpha_1} \otimes \dots \otimes \sigma_N^{\alpha_N}}$ are all the Pauli strings \rep{of the}{that constitute the} Pauli group $\mathcal{P}_N$ of $N$ qubits. One can also define the density of \rep{magic}{SRE} $m_2 = \mathcal{M}_2/N$. For pure states $ \hat \rho = \ket{\psi}\bra{\psi} $, $ \ket{\psi} $ is a stabilizer state if and only if $ \hat \rho $ has precisely $ 2^N $ nonzero components in the operator basis given by the Pauli strings, while all other components are zero~\cite{aaronson2004}. As such, the SRE provides a measure of the entropy associated with the probability distribution $ \Pi_P = \mathrm{Tr}(\hat P \hat \rho)^2 / 2^N $, adjusted by the entropy of a stabilizer state. If a mixed state can be produced by partially tracing a stabilizer pure state, \rep{it has zero magic}{its SRE is zero}. The SRE obeys the bound $ 0 \leq \mathcal{M}_2 < N \ln 2 $, where $ \mathcal{M}_2 = 0 $ if and only if the state is a stabilizer.

Computing the SRE is a costly operation, as it involves a sum over the $4^N$ elements of the Pauli group. For the permutationally-invariant system considered in this manuscript, however, its calculation can be simplified dramatically and carried out with $\mathrm{poly}(N)$ operations. In fact, two Pauli strings that have the same gates in a different order will have the same expectation values on a permutationally-invariant quantum state. This observation allows decomposing the Pauli group into independent subsets of strings that only differ by permutations. It is easy to show that there are only $\mathcal{D} = \binom{N+3}{3} = (N + 3)(N+1)(N+1)/6$ such independent sets and that the size $g$ of each set is given by the multinomial coefficient
\begin{equation}
    g = \frac{N!}{N_x!N_y!N_z!(N-N_x-N_y-N_z)!},
\end{equation}
where $N_\alpha$ ($\alpha \in \lbrace x,y,z\rbrace$) is the number of $\alpha$-gates in any of the Pauli strings of the set. Thus, for permutationally invariant states, one only has to compute $\mathcal{D} \sim N^3$ expectation values to compute the SRE, an exponential advantage compared to the general case. For further details, we point the reader to Ref.~\cite{passarelli2024magic}.

\subsection{Unravelling-independence of stabilizer entropy with negligible correlations}\label{labeloide:sec}

Our first result is that, for systems where correlations are negligible, the ensemble-average of the SRE $\mathcal{M}_2$ is \emph{unraveling-independent}. This result particularly applies to the model of Eq.~\eqref{eq:btc}, for large system sizes, as we are going to show in the following. Here we propose a theoretical argument supporting this statement for generic systems without correlations. In this subsection, we denote the ensemble averages over the quantum trajectories as $\mathbb{E}[\bullet]$, whereas for the expectation values of an observable $\hat P$ over a state $\ket{\psi^{(k)}}$ of the ensemble we use the superscript $(k)$, writing $P^{(k)}$. 

The goal is to show that $\mathbb{E}[\mathcal{M}_2]$ is unraveling-independent; to this end, we exploit the replica trick to rewrite the expectation of the logarithm in terms of the expectations of the moments of the distribution, which allows us to write
\begin{align}
    \mathbb{E}[\mathcal{M}_2] &= -\mathbb{E}\left[\log \left(\sum_{\hat P} {P^{(k)}}^4 / 2^N\right)\right] \notag \\
    &= -\lim_{n\to 0} \frac{\mathbb{E}\left[\left(\sum_{\hat P} {P^{(k)}}^4 / 2^N\right)^n\right]-1}{n}.
\end{align}
If the correlations between the different replicas are negligible, we can replace $\mathbb{E}[X^n] \approx \mathbb{E}[X]^n$ in the previous equation, where $X$ is the argument of the ensemble average in the second line, yielding the resummation
\begin{equation}
    \mathbb{E}[\mathcal{M}_2] = -\log \left(\mathbb{E}\left[\sum_{\hat P} {P^{(k)}}^4 / 2^N\right]\right).
\end{equation}
Then, after exploiting the linearity of the average, by the same reasoning one could again disregard correlations in $\mathbb{E}[P^4] \approx \mathbb{E}[P]^4$ and arrive to
\begin{equation}\label{eq:unravelings-independence}
    \mathbb{E}[\mathcal{M}_2] = -\log \left(\sum_{\hat P} {\mathbb{E}\left[P^{(k)}\right]}^4 / 2^N\right),
\end{equation}
where the average has moved all the way inside the logarithm. So, since $\hat P$ are observables, their ensemble averages are unraveling-independent, and so is $\mathbb{E}[\mathcal{M}_2]$.

\begin{figure*}[t]
    \centering
    \includegraphics[width=\textwidth]{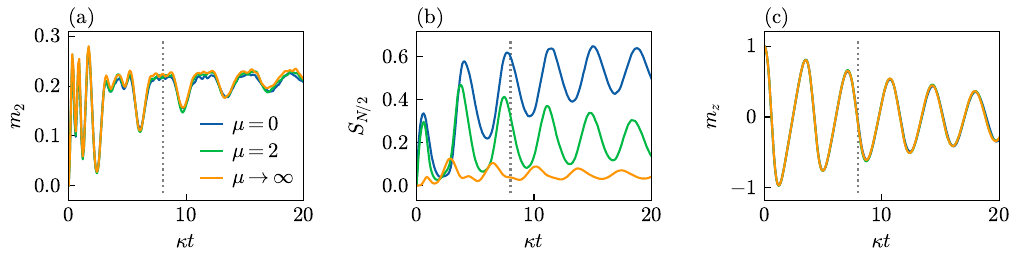}
    \caption{(a) \rep{Magic}{Nonstabilizerness} density, (b) entanglement and (c) magnetization dynamics for different unravelings ($N = 40$, $\omega_0 = 2 \kappa$, averaged over $M = 2000$ trajectories). The dotted line marks $\kappa t = 8$: the distributions of \rep{magic}{nonstabilizerness}, entanglement and magnetization of the three unravelings at this time are shown in Appendix~\ref{app:histograms}.}
    \label{fig:btc-unravelings}
\end{figure*}

\subsection{Nonstabilizerness and permutational invariance}
\label{subsec:m2-permutational-invariance}

In the case of permutationally invariant systems, Eq.~\eqref{eq:unravelings-independence} only holds only under certain conditions. The expectation value of any collective observable $ \hat O$ on a permutationally invariant quantum state scales extensively with system size, i.\,e., $\langle \hat O \rangle \sim N$. Similarly, connected correlations of collective observables are also extensive, including the variance of $\hat O$: $\text{var}(O) = \langle \hat O^2 \rangle - {\langle \hat O \rangle}^2 \sim N$, which implies $\delta O/O \sim 1/\sqrt{N}$, where $\delta O = \sqrt{\text{var}(O)}$. Therefore, correlations become negligible in the large-$N$ limit, justifying the application of Eq.~\eqref{eq:unravelings-independence}.

Thus, the argument above is rigorously valid for permutationally invariant states only in the thermodynamic limit $N \to \infty$, when one recovers Eq.~\eqref{eq:sre-density} with $\mathrm{Tr}(\hat \rho^2) = 1$. For finite $N$, this remains an approximation that heuristically holds when $N$ is large enough for correlations to be disregarded. The crucial insight from this analysis is that the stabilizer entropy is unravelling-independent in our setup because it is a nonlinear function of \emph{expectation values} of Pauli strings over the ensemble states, rather than a nonlinear function of the quantum states themselves.

Other measures of quantum complexity, however, might be more sensitive than $\mathcal{M}_2$ to residual correlations at finite size, leading to a stronger dependence on the monitoring protocol. A notable example is the bipartite entanglement entropy, which, for a pure state, is an entanglement monotone defined as~\cite{nielsen2010}
\begin{equation}
    S_{N/2} = -\mathrm{Tr}(\hat \rho_{N/2} \log \hat \rho_{N/2}),
\end{equation}
where $\hat \rho_{N/2}$ is the reduced density matrix after partial tracing the degrees of freedom of half the system from the state $\hat \rho = \ket{\psi}\bra{\psi}$.

Ref.~\cite{passarelli2023} analyzed the entanglement properties of the superradiant model in Eq.~\eqref{eq:btc} across its phase diagram, showing that the entanglement entropy shows significant differences between unravelings, particularly in its scaling behavior in the low-dissipative regime. In the QJ unraveling, it transitions from an area law\add{, where the enetanglement entropy is independent of $N$,} to a $\log(\log N$) scaling, with a $\log N$ peak at a critical point. In contrast, in the QSD unraveling, it undergoes an area-to-area transition with a logarithmic peak at the same critical point \add{(so, the entanglement entropy is size-independent across the phase diagram except at the critical point)}. This stark difference arises because entanglement entropy is a nonlinear function of the quantum state, making it intrinsically sensitive to correlations, whereas stabilizer entropy, being expressible solely in terms of expectation values, is not.

To illustrate this point, we show one example of dynamics in Fig.~\ref{fig:btc-unravelings} for a system of $N = 40$ spins, in the BTC phase [$\omega_0 = 2 \kappa$ in Eq.~\eqref{eq:btc}]. The Lindblad dynamics is unraveled in terms of pure-state quantum trajectories, and averages are computed over $M = 2000$ realizations for all considered unravelings, specified by the value of the parameter $\mu$ in Eq.~\eqref{eq:unravelings}. In panel~(a), where we plot the nonstabilizerness density, the curves relative to the different unravelings are very close to each other (we have numerically verified that the difference between curves decreases by increasing $N$), whereas in panel~(b) we see that the unravelings have completely different average bipartite entanglement. As expected, the magnetization along the $z$ axis, defined as $m_z = \langle \hat S_z \rangle / S $, is the same among all unravelings, panel~(c), as for any other observable. Similar results are valid for the full distributions along quantum trajectories, as we show in Appendix~\ref{app:histograms}.

\rep{These findings highlight that different measures of quantum complexity capture distinct physical aspects of monitored systems.}{These findings highlight that, in a regime where quantum features are present as testified by the nonzero entanglement entropy, different physical quantities relevant for quantum information behave differently under monitored dynamics.} While the stabilizer entropy remains largely unraveling-independent due to its reliance on expectation values, the entanglement entropy---being inherently sensitive to correlations---exhibits a strong dependence on the monitoring protocol. This contrast highlights the fact that quantum complexity manifests in fundamentally different ways depending on the chosen metric, as each assigns a different weight to correlations in the system.

\section{Nonstabilizerness in the mean-field limit}\label{sec:mf}

In the thermodynamic limit, the \add{collective magnetizations of the superradiant model of Eq.~\eqref{eq:btc} behave like those of a classical system: The correlations between collective spin variables scale more slowly with the system size $N$ than their expectations. So, when $N\to \infty$, the correlations become negligible compared to the expectations and the latter ones are} exactly described by mean-field theory. This means that, in this limit, for the purpose of evaluating the expectation values of system observables, the system state can be approximated as a product state of single-particle density matrices, 
\begin{equation}\label{haro:eqn}
      \hat \rho = \bigotimes_k \hat \rho_k\,,
\end{equation}
 with
\begin{equation}
    \hat \rho_k = \frac{\hat \sigma_0}{2} + \frac{1}{2} \vec{m} \cdot \vec{\hat \sigma},
\end{equation}
where $\vec{m} = (m_x, m_y, m_z)$. Also notice that, since ${\lvert \vec{m} \rvert}^2 = 1$ in the maximum spin subspace, the system is in a pure state, as its purity reads $\Tr(\hat \rho_k^2) = (1 + {\lvert \vec{m} \rvert}^2)/2 = 1$. 

We are going to see in Sec.~\ref{sec:finite-size} that the Lindblad average density matrix has a purity smaller than one in the BTC phase, and this fact persists in the large-size limit. Therefore in this limit the state is not pure and for what we have just discussed the factorized form Eq.~\eqref{haro:eqn} is not exact. It is a good approximation for the local observables (for which the mean field is exact in the thermodynamic limit), but the correlations affect the behavior of the nonstabilizerness, as we are going to see.

The key point is that, since correlations in collective observables are negligible in the thermodynamic limit, one is allowed to express the density matrix as a product state of individual components, when one deals with evaluation of expectations of observables. We will see in the following that this assumption may lead to incorrect predictions for the \rep{magic}{nonstabilizerness} at finite sizes in some regions of the phase diagram, where the steady state is actually mixed. Indeed, the mean-field approximation is not well defined for mixed states, and this will have important consequences on the evaluation of quantum complexity measures.

For $N\to\infty$, the system can be entirely described by the equations of motion of the expectation values of the magnetization components, $m_\alpha = \langle \hat S_\alpha \rangle / S $. These read
\begin{equation}\label{eq:mag}
    \begin{cases}
        \partial_t m_x(t) = \kappa \, m_x(t) \, m_z(t);\\
        \partial_t m_y(t) = -\omega_0 m_z(t) + \kappa \, m_y(t) \, m_z(t); \\
        \partial_t m_z(t) = \omega_0 m_y(t) - \kappa \, [m_x^2(t) + m_y^2(t)].
    \end{cases}
\end{equation}
Depending on the ratio $\Omega = \omega_0 / \kappa$, MF predicts that the system can be in one of two different phases, which can be found by looking for the stationary solutions of the dynamical system, $\partial_t \vec{m} = 0$. If $\Omega < 1 $, one gets the two possible solutions
\begin{equation}\label{eq:fp-mag}
    \overline{m_x} = 0, \quad \overline{m_y} = \Omega, \quad \overline{m_z} = \pm\sqrt{1-\Omega^2},
\end{equation}
where the overline indicates the fixed point. Since $\overline{m_z} \ne 0$, this is a magnetized phase. Instead, if $\Omega \ge 1$, one gets the BTC phase, characterized by the fixed point 
\begin{equation}\label{eq:fp-btc}
    \overline{m_x} = \pm\sqrt{1-\Omega^{-2}}, \quad \overline{m_y} = \Omega^{-1}, \quad \overline{m_z} = 0.
\end{equation}
When dissipation dominates, $\Omega = \omega_0 / \kappa < 1$, the system evolves towards a steady state with a finite value of the magnetization $\overline{m_z}$. On the other hand, if 
$\Omega > 1$, the fixed point of the dynamics is an oscillation center, and there are no steady states: instead, the system is found in the boundary time crystal phase, displaying persistent oscillations of the magnetization around its fixed point. The critical point $\Omega_c = 1$ marks a second-order dissipative phase transitions between these two phases.

Using Eq.~\eqref{eq:sre-density} and the fact that the state is a pure product state for $N\to\infty$, it is easy to compute the \rep{magic density}{density of SRE} of the average state $\hat \rho$ during its dynamics. In fact, we have $\mathcal{M}_2(\hat \rho) = N \mathcal{M}_2(\hat \rho_k)$, thus $\mathcal{M}_2(\hat \rho) / N = m_2(\hat \rho) = \mathcal{M}_2(\hat \rho_k)$ and the latter is given by
\begin{equation}\label{eq:m2-product-state}
    m_2(\hat \rho) = -\log \left( \frac{1+m_x^4 + m_y^4 + m_z^4}{2}\right),
\end{equation}
where the magnetization components obey Eqs.~\eqref{eq:mag}. We note that, in the thermodynamic limit, the density of \rep{magic}{SRE} is always finite, thus the stabilizer entropy always scales extensively with $N$, except when $m_x^4 + m_y^4 + m_z^4 = 1$. This only occurs when one of the magnetization components is equal to $\pm 1$ while the rest are zero, i.\,e., when the state is a stabilizer and is fully polarized along one of the three coordinate axes. We remark that Eq.~\eqref{eq:m2-product-state} is only valid because correlations are negligible.

\begin{figure}[t]
    \centering
    \includegraphics[width=\columnwidth]{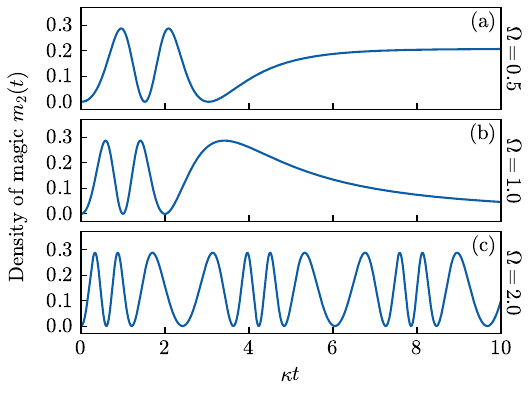}
    \caption{Density of (mean-field) \rep{magic}{nonstabilizerness} as a function of time, obtained from Eqs.~\eqref{eq:mag} and~\eqref{eq:m2-product-state}. (a)~$\Omega = \omega_0/\kappa = 0.5$. (b)~$\Omega = 1$. (c)~$\Omega = 2$. The initial condition is $\vec{m}(0) = (0,0,1)$.}
    \label{fig:btc-mf-dyn}
\end{figure}

Examples of \rep{magic}{SRE} dynamics in the thermodynamic limit are shown in Fig.~\ref{fig:btc-mf-dyn} for different choices of the system parameters. The spin system is initialized in the fully polarized state $\vec{m}(0) = (0,0,1)$. We numerically integrate Eqs.~\eqref{eq:mag} using implicit backward differentiation formulas of order 5 (BDF). Comparing the panels, we see that when $\Omega \le \Omega_c$ the \rep{magic}{SRE} density quickly saturates to its stationary value after a transient time that depends on the initial state; on the other hand, for $\Omega > \Omega_c$, it displays never-ending oscillations akin to the magnetization oscillations in the BTC phase.

\begin{figure}[t]
    \centering
    \includegraphics[width = \columnwidth]{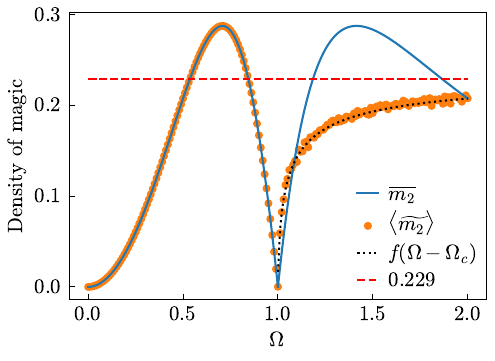}
    \caption{Density of \rep{magic}{nonstabilizerness} of the fixed point [$\overline{m_2}$, Eq.~\eqref{eq:m2-mf}] and average density of \rep{magic}{nonstabilizerness} along the closed orbits around the fixed point [$\langle \widetilde{m_2} \rangle$, Eq.~\eqref{eq:m2-avg}] in the mean-field limit ($\Omega = \omega_0 / \kappa$). The black dotted line is the numerical fit of Eq.~\eqref{eq:m2-fit}. The red dashed line indicates the saturation value, estimated using Eq.~\eqref{eq:m2-estimate}.}
    \label{fig:btc-mf}
\end{figure}

Interestingly, the \rep{magic density}{density of SRE} of the fixed point can be computed analytically using Eqs.~\eqref{eq:fp-mag} and~\eqref{eq:fp-btc}. It reads
\begin{equation}\label{eq:m2-mf}
    \overline{m_2} = 
    \begin{cases}
    \displaystyle-\log(\Omega^4-\Omega^2+1) &\text{if $\Omega < 1$;}\\
    \displaystyle-\log\left(\Omega^{-4}-\Omega^{-2}+1\right) &\text{if $\Omega \ge 1$.}
    \end{cases}
\end{equation}
It is shown in Fig.~\ref{fig:btc-mf}, blue line. At the critical point $\Omega = \Omega_c = 1$, the fixed-point \rep{magic}{nonstabilizerness} density is zero. This is because the fixed point in that case is the fully polarized eigenstate of $\hat S_y$, a stabilizer state. Moreover, the derivative of the \rep{magic}{nonstabilizerness} density is singular at $\Omega = \Omega_c$, as $m_2$ has a cusp there, and has a finite jump $\Delta \overline{m_2}' = 4$. Thus, even if \rep{magic}{nonstabilizerness} does not display any change of behavior between the magnetized and the time-crystal phase, it still shows singular behavior at criticality.

The \rep{magic}{SRE} density is zero also for $\Omega = 0$, where the fixed point is a steady state and is the dark state $\ket{\psi_D} = \ket{\downarrow\downarrow\cdots\downarrow}$ (\add{i.\,e., a pure state such that $\hat L \ket{\psi_D} = 0$,} also a stabilizer state), and $\Omega \to \infty$, where fixed points are the eigenstates of $\hat H = \omega_0 \hat S_x$. The maximum \rep{magic}{SRE} density is found for $\Omega^*_1 = 1/\sqrt{2}$ and $\Omega^*_2 = \sqrt{2}$ and is equal to $\overline{m_2}^* = 2\log 2 - \log 3$, which coincides with the \rep{magic}{stabilizer entropy} of the $T$ state $\ket{T} \sim \ket{\uparrow} + \exp(i \pi / 4) \ket{\downarrow}$.

However, it is important to note that, in the BTC phase, the fixed point of the classical MF dynamics is not a steady state, but an oscillation center in phase space~\cite{iemini2018}. Thus, in this phase it makes more sense to look at the average \rep{magic}{nonstabilizerness} along the closed orbits encircling the fixed point, rather than the fixed point itself. 

On the one hand, by averaging over many multiples of the oscillation period, along the orbit of the many-body limit cycle, one can extract the fixed-point magnetization components from the time traces. In fact, for any collective observable, averaging over limit-cycle orbits gives back its value at the fixed point. 

On the other hand, this is not necessarily true for for \rep{magic}{the SRE}. In this case, the average over the orbit of the limit cycle is \emph{not} equivalent to the \rep{magic}{SRE} of the fixed point and is generally dependent on the initial condition, which determines the orbit in the semiclassical phase space. 

To evaluate $m_2$ along these orbits, we proceed numerically. First, we generate $N_\text{avg} = 1000$ initial conditions of the form $\vec{m}(0) = (\sin \theta \cos \phi, \sin \theta \sin \phi, \cos \theta)$, where $\theta \in [0, \pi]$ and $\phi \in [0, 2\pi]$ are random variables uniformly distributed in their domains. For each of them, denoted $k$, we solve Eqs.~\eqref{eq:mag} up to a long time $\tau$ and compute the instantaneous \rep{magic}{nonstabilizerness} density $m_2^{(k)}(t) $ of this classical trajectory using Eq.~\eqref{eq:m2-mf}. In our simulation, we consider $\kappa \tau = 1000$. Then, we consider the (long-time) average \rep{magic}{SRE} along the trajectory,
\begin{equation}\label{eq:m2-lta}
    \widetilde{m_2}^{(k)} = \frac{2}{\tau} \int_{\tau/2}^\tau m_2^{(k)}(t) \, dt,
\end{equation}
and finally average this quantity over the $N_\text{avg}$ initial states:
\begin{equation}\label{eq:m2-avg}
    \langle \widetilde{m_2} \rangle = \frac{1}{N_\text{avg}} \sum_{k=1}^{N_\text{avg}} \widetilde{m_2}^{(k)}.
\end{equation}
The result of this procedure is shown in Fig.~\ref{fig:btc-mf} with orange scatter points. Error bars (denoting the standard error, i.\,e., the standard deviation divided by $\sqrt{N_\text{avg}}$) are not visible as they are smaller than the point size. We see that, in the magnetized phase $\Omega < 1$, the \rep{magic}{stabilizer entropy} computed by averaging over random initial states coincides with the \rep{magic}{SRE} of the stationary state, whereas for $\Omega > 1$, in the BTC phase, the two quantities differ from each other. In either case, the \rep{magic}{nonstabilizerness} density is finite, so the system always has extensive \rep{magic}{SRE} and there are no \rep{magic}{nonstabilizerness} phase transitions associated with the mean-field transition. Yet, the cusp at criticality remains and even becomes more pronounced, as the right derivative of the averaged \rep{magic}{SRE} density at $\Omega_c$ becomes larger than the corresponding derivative of the fixed-point \rep{magic}{stabilizer entropy}. Instead, the maximum at $\Omega_2^* = \sqrt{2}$ is gone, leaving the room to a monotonic growth of $\langle \widetilde{m_2}\rangle $ with the driving frequency. 

A theoretical argument allows to compute its saturation value. When $\Omega \to \infty$, the oscillation frequency of the magnetization components goes to infinity as well. Thus, the three components will uniformly sample their respective domain and an average over time and initial conditions will be equivalent to an average over the solid angle of the \rep{magic}{SRE} density of the spherical vector $\vec{m} = (\sin \theta \cos \phi, \sin \theta \sin \phi, \cos \theta)$. Thus, one can estimate
\begin{align}\label{eq:m2-estimate}
    \langle \widetilde{m_2} \rangle^{(\Omega\to\infty)} &\approx -\frac{1}{4\pi} \int_{0}^{\pi} \sin \theta d\theta \int_0^{2\pi} d \phi\notag \\&\quad\log \left[\frac{1 + \cos^4\theta + \sin^4 \theta (\sin^4 \phi +  \cos^4 \phi)}{2}\right] \notag \\ & \approx 0.229.
\end{align}
This value is indicated in Fig.~\ref{fig:btc-mf} with a red dashed line. One can fit the numerical data for $\Omega > 1$ with the generalized Michaelis-Menten law~\footnote{While this empyrical law holds no physical meaning in this analysis, it is commonly used to describe saturating behaviors that start as power laws and are not exponential, and to extract the saturation value. We tested other empirical laws, all of them gave comparable estimates for the saturation value.}
\begin{equation}\label{eq:m2-fit}
    f(\Omega -\Omega_c) = \frac{m_2^\text{(sat)} {(\Omega - \Omega_c)}^\alpha}{a + {(\Omega - \Omega_c)}^\alpha}.
\end{equation}
In this way, one can estimate $m_2^\text{(sat)} \approx 0.232$, $\alpha \approx 0.77$, and $a \approx 0.11$. The value of the critical exponent $\alpha$ suggests that the derivative of $m_2$ is divergent at criticality~\footnote{Even though numerical fits suggest the derivative may diverge at criticality, they are not entirely conclusive since the numerical values are affected by finite errors, which mask potential divergences.}. This fit is shown in Fig.~\ref{fig:btc-mf} with a black dotted line. Note that the fitted saturation value, $m_2^\text{(sat)}$, is very close to our other estimate, Eq.~\eqref{eq:m2-estimate}.

\section{Finite-size results}
\label{sec:finite-size}

\begin{figure}[t]
    \centering
    \includegraphics[width = \columnwidth]{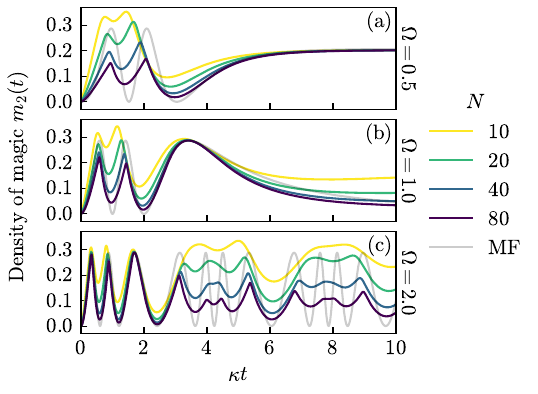}
    \caption{Density of \rep{magic}{nonstabilizerness} of the average state evolving according to the Lindblad equation [Eq.~\eqref{eq:btc}] as function of time, for several system sizes $N$, compared with the mean-field result. (a)~$\Omega = \omega_0/\kappa = 0.5$. (b)~$\Omega = 1$. (c)~$\Omega = 2$. The initial condition is the fully polarized state $\ket{\psi(0)} = \ket{\uparrow\uparrow \dots \uparrow}$.}
    \label{fig:btc-finitesize-dyn}
\end{figure}

\subsection{Nonstabilizerness of the average state}

We now turn our attention to the study of nonstabilizerness in finite-size systems. We start by directly solving the unconditioned Lindblad dynamics of Eq.~\eqref{eq:btc}, and we compute $\mathcal{M}_2$ of the average state according to Eq.~\eqref{eq:sre-density}. We integrate the Lindblad equation using the 4th order Runge-Kutta method. Due to the strong symmetry of the model, we can represent the entire dynamics in the maximum spin subspace ($S = N/2$), so the density matrix has a linear dimension $D = N + 1$~\cite{shammah2018}.

Our results are shown in Fig.~\ref{fig:btc-finitesize-dyn}, for system sizes up to $N = 80$, for the same values of $\Omega$ shown in Fig.~\ref{fig:btc-mf-dyn}. In all panels, we see that the density of \rep{magic}{SRE} of the average state depends on the system size.  For $\Omega \le 1$, we see that the long-time density of \rep{magic}{SRE} of the average state coincides with the density of \rep{magic}{stabilizer entropy} calculated using the mean-field approximation (the gray line), whereas for $\Omega > 1$ this is not the case. In fact, in this regime the density of \rep{magic}{SRE} \emph{decreases} with $N$ at long times, in contrast with the finite $m_2$ observed in MF dynamics. 

This apparent contradiction is due to the fact that finite-size dynamics for large $N$ only converge to the MF solution for $\Omega \le 1$. Instead, in the BTC phase $\Omega > 1$, the infinite-time limit and the thermodynamic limit do not commute:
\begin{equation}\label{eq:limits}
    \lim_{t\to\infty} \lim_{N\to\infty} \ne \lim_{N\to\infty} \lim_{t\to\infty}.
\end{equation}
Let us consider the left-hand side of Eq.~\eqref{eq:limits}. If one takes the thermodynamic limit first, then the dynamics of collective observables is described by Eqs.~\eqref{eq:mag}, the system state is always pure, and the behavior of \rep{magic}{the stabilizer entropy} is the one shown in Fig.~\ref{fig:btc-mf-dyn}. 

\begin{figure}[t]
    \centering
    \includegraphics[width = \columnwidth]{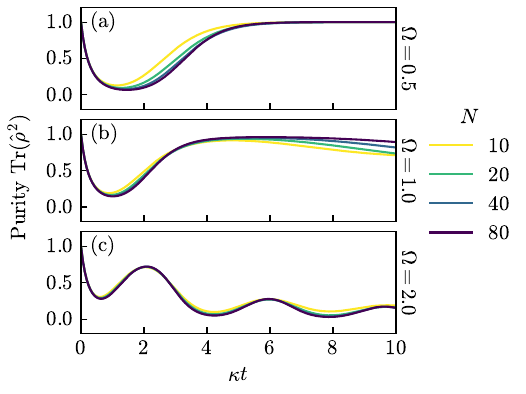}
    \caption{Purity of the average state evolving according to the Lindblad equation [Eq.~\eqref{eq:btc}] as function of time, for several system sizes $N$. (a)~$\Omega = \omega_0/\kappa = 0.5$. (b)~$\Omega = 1$. (c)~$\Omega = 2$. The initial condition is the fully polarized state $\ket{\psi(0)} = \ket{\uparrow\uparrow \dots \uparrow}$.}
    \label{fig:btc-purity-dyn}
\end{figure}

If instead one takes the long-time limit first [i.\,e., the right-hand side of Eq.~\eqref{eq:limits}], the situation is much different. In the BTC phase, for any finite $N$ the Lindblad dynamics converges to a unique mixed steady state, regardless of the initial state. In this mixed quantum state, correlations are not negligible, Eq.~\eqref{haro:eqn} is just an approximation that describes well the expectation of observables, and the MF picture painted in Sec.~\ref{sec:mf} breaks down.

The emergence of the time-crystal phase in the weakly dissipative regime is signaled by the fact that the amplitude of the magnetization oscillations decays more slowly as $N$ increases, but they are only persistent for $N\to\infty$: For any finite size, they will eventually decay, yielding a mixed steady state with $\lvert \vec{m}\rvert < 1$ and a density of \rep{magic}{SRE} that goes to zero as $N$ grows. 

Both these facts are not consistent with the mean-field limit, where $\hat \rho$ is a pure state with a finite density of \rep{magic}{SRE}. In particular, see Fig.~\ref{fig:btc-purity-dyn}, where we plot the purity of the average state for the same parameters as Fig.~\ref{fig:btc-finitesize-dyn}. Here is evident that, for any finite size, the purity of the average state is smaller than one in the BTC phase, as a consequence of the noncommutativity of the thermodynamic limit and the long-time limit, Eq.~\eqref{eq:limits}, and of the existence of residual correlations in the state.

Instead, the two limits in Eq.~\eqref{eq:limits} do commute in the magnetized phase. Here, the purity goes to one, and the steady-state \rep{magic}{nonstabilizerness} converges to its MF limit.

\begin{figure}[t]
    \centering
    \includegraphics[width = \columnwidth]{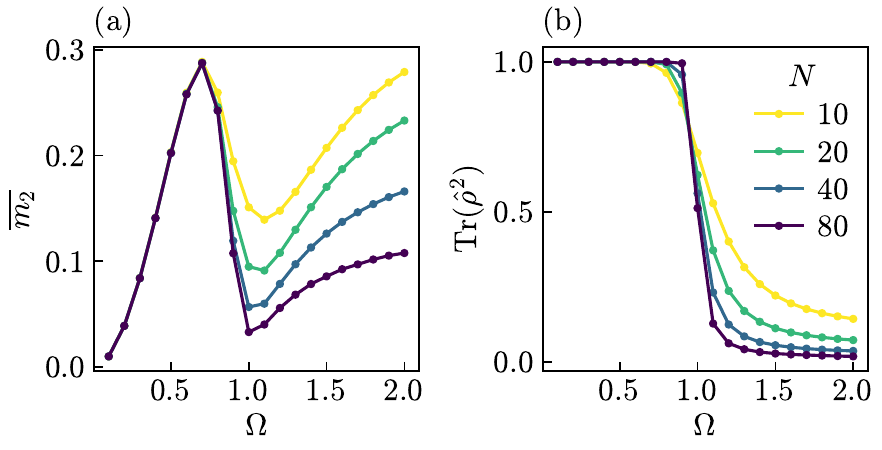}
    \caption{(a) Finite-size density of \rep{magic}{nonstabilizerness} of the steady state for different system sizes, in a deterministic Lindblad evolution, as a function of $\Omega = \omega_0 / \kappa$. (b) Purity of the steady state as a function of $\Omega$.}
    \label{fig:btc-finite}
\end{figure}

The behavior of the density of \rep{magic}{SRE} of the steady state is shown in Fig.~\ref{fig:btc-finite}(a) as a function of the driving frequency, further confirming that the Lindblad equation does not agree with the MF prediction of a finite density of \rep{magic}{stabilizer entropy} for large $N$. This mismatch is accompanied by a sudden decrease in the purity of the steady state, as shown in Fig.~\ref{fig:btc-finite}(b), confirming the important role of correlations in this region for finite sizes. As long as the steady state is pure ($\Omega \le 1$), the MF predictions and the \rep{magic}{nonstabilizerness} of the finite-size state evolving through the Lindblad equation [Eq.~\eqref{eq:btc}] agree with each other (compare this with the \rep{magic}{SRE} of the fixed point shown in Fig.~\ref{fig:btc-mf} for $\Omega \le 1$). This is no longer the case when $\text{Tr}\,\hat\rho^2 <1$.

For this reason, we now turn our attention to the conditional monitored dynamics given by the unravelings described in Sec.~\ref{sec:model}.

\subsection{Nonstabilizerness of unravelings}

\begin{figure*}[t]
    \centering
    \includegraphics[width = \linewidth]{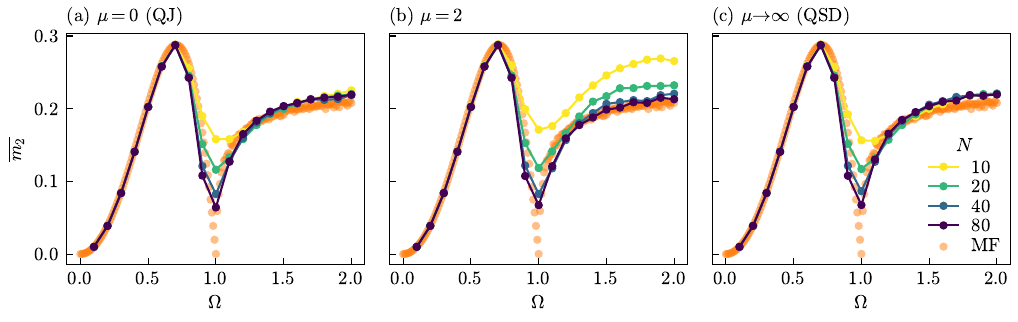}
    \caption{Finite-size long-time averaged \rep{magic}{nonstabilizerness} density for different system sizes in different unravelings, see Eq.~\eqref{eq:unravelings}. (a) Quantum Jump ($\mu = 0$). (b) $\mu = 2$. (b) Quantum State Diffusion ($\mu \to \infty$). We also show the average mean-field \rep{magic}{SRE} $\langle \widetilde{m_2} \rangle $, Eq.~\eqref{eq:m2-avg} (orange points).}
    \label{fig:btc-qj-mu2-qsd}
\end{figure*}

In this section, we show that the inconsistencies pointed out in the previous section disappear if, instead of computing the \rep{magic}{nonstabilizerness} of the average state, one considers the average \rep{magic}{stabilizer entropy} over ensembles of monitored quantum trajectories. 

Indeed, in Fig.~\ref{fig:btc-qj-mu2-qsd} we report the average long-time \rep{magic}{SRE} of ensembles of $M = 500$ quantum trajectories, in three unravelings. In the three panels, orange scatter points represent the mean-field results discussed in Sec.~\ref{sec:mf}. For $\Omega<1$ they correspond to the steady state, while for $\Omega >1$ to the average over time and random initial conditions. First of all, we see that the \rep{magic}{SRE} density in the $\Omega < 1$ phase is essentially unchanged compared to Fig.~\ref{fig:btc-finite}, meaning that each quantum trajectory of the ensemble behaves very closely to the average state in this phase. This is also consistent with the early findings, showing that the variances of collective observables over the ensemble in this phase go to zero for large $N$~\cite{passarelli2023}. 

The situation is much different from the average one in the $\Omega > 1$ phase, where we observe a finite, size-independent, density of \rep{magic}{SRE} that tells us that the \rep{magic}{stabilizer entropy} of the ensemble trajectories is extensive, as opposed to the \rep{magic}{nonstabilizerness} of the average state $\hat \rho$. Moreover, we see that the ensemble average of $m_2$, where each trajectory starts from the same initial state but follows a stochastic dynamics due to the random measurements performed by the environment, is essentially identical to the mean-field \rep{magic}{SRE} averaged over many initial states, Eq.~\eqref{eq:m2-avg} (see the orange points in Fig.~\ref{fig:btc-qj-mu2-qsd}). This observation tells us that the intrinsic randomness of quantum trajectories allows the system to explore its phase space, and that this information is completely hidden at the level of the average state evolving according to the deterministic Lindblad equation.

Moreover, we observe that the \rep{magic}{nonstabilizerness} in all the unravelings is exactly the same, when the system size is sufficiently large, as we have analytically shown in Sec.~\ref{labeloide:sec}, in contrast with the behavior of the bipartite entanglement entropy~\cite{passarelli2023}. %
We see that all the unravelings give the same \rep{magic}{stabilizer entropy} except for finite-size corrections at small sizes, as shown in Fig.~\ref{fig:btc-qj-mu2-qsd}(b--c) for $\mu = 2$ and $\mu \to \infty$.

\begin{figure}[b]
    \centering
    \includegraphics[width = \columnwidth]{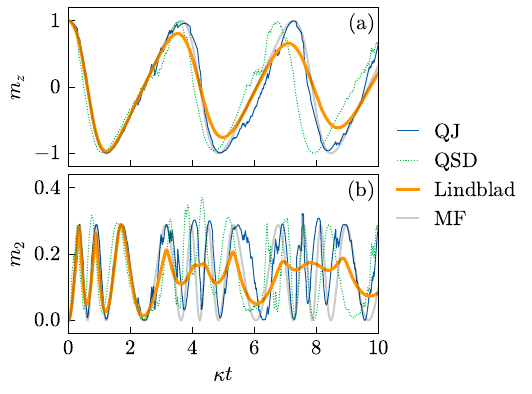}
    \caption{(a) Magnetization dynamics along single trajectories of the quantum jump and quantum state diffusion unravelings, versus the dynamics of the average state and its mean-field limit ($\Omega = 2$). (b) Same as panel (a) but for the density of nonstabilizerness $m_2$. Finite-size dynamics are for $N = 40$. The initial condition is the fully polarized state $\ket{\psi(0)} = \ket{\uparrow\uparrow \dots \uparrow}$, with $\vec{m}(0) = (0,0,1)$.}
    \label{fig:btc-onetraj-dyn}
\end{figure}

All these numerical results are consistent with the interpretation discussed in Sec.~\ref{subsec:m2-permutational-invariance}: for large $N$, the SRE is unraveling-independent because correlations become negligible when computing the expectation values of collective observables. Indeed we can see by comparing the different panels of Fig.~\ref{fig:btc-qj-mu2-qsd} that for small $N$ the \rep{magic}{nonstabilizerness} density is \emph{not} unraveling independent, due to correlations, and that unraveling-independence only kicks in for $N \gtrsim 20$ (empirically). %

\begin{figure*}[t]
	\centering
	\includegraphics[width=\textwidth]{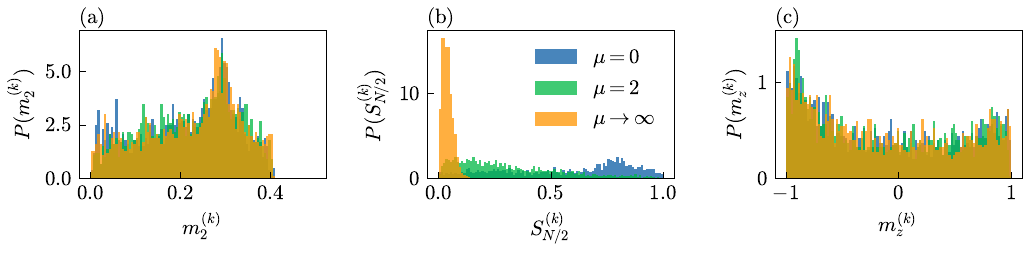}
	\caption{Histograms showing the probability density of the (a) \rep{magic}{nonstabilizerness} density, (b) entanglement entropy and (c) magnetization in the same ensembles as Fig.~\ref{fig:btc-unravelings}. Data are binned in 100 bins. Parameters: $N = 40$, $\Omega = 2$, $\kappa t = 8$.}
	\label{fig:btc-hist}
\end{figure*}

Thus, unraveling the dissipative dynamics in terms of quantum trajectories is able to resolve the inconsistency between finite-size simulations and the thermodynamic limit in the BTC phase, for two reasons. First of all, the states of the ensemble of quantum trajectories are all pure by construction, therefore the SRE is a proper complexity measure for each of them and there are no issues associated with nonpure states like for the \rep{magic}{stabilizer entropy} of the Lindblad steady state in the BTC phase. %
Second, due to the intrinsic randomness of quantum trajectories, the dynamics of the single trajectories is much different from the dynamics of the Lindblad average state, and averaging over trajectories and time one gets a result well described in the large-size limit by the average over time and random initial states of MF trajectories.

Even though we do not have an analytical explanation of this fact, we can gain some insight by comparing stochastic trajectories and a mean-field one. %
For instance, in Fig.~\ref{fig:btc-onetraj-dyn}(a), we report the dynamics of $m_z$ along a single quantum trajectory of the quantum jump and quantum state diffusion unravelings in the BTC phase $\Omega = 2$, compared with the corresponding Lindblad evolution, for $N = 40$. Here we see a crucial difference between the dynamics along a single trajectory and the average state. Along each quantum trajectory, we observe essentially persistent (non-periodic) oscillations of the magnetization even at finite size, i.\,e, the amplitude does not monotonically decay. However, different trajectories have slightly displaced oscillations due to randomness that, once averaged, yield a magnetization dynamics whose oscillations decay over time---the Lindblad evolution. From this point of view, single trajectories in the BTC phase resemble the behavior of the system in the thermodynamic limit, since for each of them the fixed point of the dynamics is not a steady state.

The same considerations apply to the density of \rep{magic}{SRE} as well, as shown in Fig.~\ref{fig:btc-onetraj-dyn}(b) for the same parameters. Also in this case, the \rep{magic}{SRE} dynamics along single trajectories more closely resembles the dynamics in the MF limit, compared to the Lindblad \rep{magic}{nonstabilizerness} where instead oscillations decay over time and \rep{magic}{the stabilizer entropy} goes to zero. Moreover, since the SRE is a nonlinear function, its ensemble average over quantum trajectories does not coincide with the SRE of the average state. Therefore we think it is reasonable that its value, long-time averaged and also averaged over the pure states of the ensemble, would resemble the MF limit (averaged over many initial states) more than the \rep{magic}{nonstabilizerness} of the average state for finite $N$.

\section{Conclusions}
\label{sec:conclusions}

In this paper, we have highlighted the fundamentally different ways in which nonstabilizerness and entanglement entropy quantify quantum complexity in open many-body systems, focusing on the case of a boundary time crystal. While both are nonlinear functions of the quantum state, they respond to the monitored dynamics in markedly distinct ways. Entanglement entropy is highly sensitive to the choice of the unraveling protocol. In contrast, we have shown that the stabilizer entropy becomes largely unraveling-independent in the thermodynamic limit, despite being a nonlinear quantity. This robustness stems from the permutational symmetry of the model, which suppresses correlations and makes the stabilizer entropy dependent on expectation values of collective observables only.

At the critical point of the measurement-induced phase transition, the two complexity measures reveal complementary aspects of the behavior of the system. The entanglement entropy exhibits a genuine phase transition with unraveling-dependent scaling, while \rep{magic}{nonstabilizerness} displays a cusp singularity in its derivative while, importantly, remaining extensive in system size across both phases. This result suggests that the superradiant model hosts highly nonclassical resources, even in the absence of entanglement.

Our results also highlight a crucial limitation of mean-field theory in capturing nonstabilizerness. While the mean-field description correctly predicts the behavior of \rep{magic}{nonstabilizerness} in the magnetized phase, it fails in the BTC phase due to the mixed nature of the unconditional steady state. %
We have shown that quantum trajectories provide an alternative framework to describe \rep{magic}{nonstabilizerness} in this regime, revealing its extensive scaling with system size and its independence from the unraveling scheme.

These findings suggest that boundary time crystals provide a promising platform for exploring the role of nonstabilizerness in open quantum systems, beyond conventional entanglement-based approaches. Future investigations could address how these results extend to models without permutational symmetry, or how other measures of nonstabilizerness behave in monitored dynamics, offering general insights into the emergence of quantum complexity in the presence of dissipation and measurement.

\begin{acknowledgments}
    We thank R. Fazio for helpful discussions. G.\,P. and A.\,R. acknowledge financial support from PNRR MUR Project PE0000023-NQSTI. We acknowledge computational resources from the CINECA award under the ISCRA initiative, and from MUR, PON “Ricerca e Innovazione 2014-2020”, under Grant No.~PIR01\_00011 - (I.Bi.S.Co.). This work was supported by PNRR MUR project~PE0000023 - NQSTI, by the European Union’s Horizon 2020 research and innovation programme under Grant Agreement No~101017733, by the MUR project~CN\_00000013-ICSC (P.\,L.),  and by the  QuantERA II Programme STAQS project that has received funding from the European Union’s Horizon 2020 research and innovation programme under Grant Agreement No~101017733 (P.\,L.).
\end{acknowledgments}

\appendix

\section{\rep{Magic}{Nonstabilizerness}, entanglement and magnetization distributions across the unravellings}
\label{app:histograms}

To gain further insights into the observation that the stabilizer entropy is unraveling-independent while the entanglement entropy is not, we also studied the \rep{magic}{stabilizer entropy} distributions on the different ensembles of quantum trajectories shown in Fig.~\ref{fig:btc-unravelings}. For instance, in Fig.~\ref{fig:btc-hist}(a) we show the histogram of the density of \rep{magic}{SRE} of the $M = 2000$ trajectories of the three ensembles, at $\kappa t = 8$ (marked in the panels of Fig.~\ref{fig:btc-unravelings} with vertical dotted lines). The histograms represent the probability density of a certain value of the \rep{magic}{nonstabilizerness} density $m_2^{(k)}$ on the state $\ket{\psi_k}$ of the unraveling considered. Besides having a similar mean, the distributions look identical. This is in line with the behavior of the magnetization, see Fig.~\ref{fig:btc-hist}(c), and completely different from the entanglement entropy, Fig.~\ref{fig:btc-hist}(b). These observations equally apply to other times $\kappa t$.

\end{document}